\documentclass[aps,prb,reprint,superscriptaddress]{revtex4-2}

\usepackage{graphicx}
\usepackage{bbm}
\usepackage{braket}
\usepackage{amsmath}
\usepackage{amsfonts}
\usepackage{graphicx}
\usepackage{amssymb}
\usepackage{color}
\usepackage{hyperref}
\usepackage{listings}
\usepackage{array}
\usepackage{physics}
\renewcommand{\vr} {{\bf r}}
\def\sss{\scriptscriptstyle\rm}
\newcommand{\uGLLB}{^{\sss 2DGLLB}}
\newcommand{\uGLLBpSCp}{^{\sss 2DGLLB(-SC)}}
\newcommand{\uTW}{^{\sss 2D}}
\newcommand{\uLDA}{^{\sss 2DLDA}}
\newcommand{\uTWBK}{^{\sss 2DB88}}
\newcommand{\uTWBKpLDAp}{ ^{ \sss 2DB88(-2DLDA)}}

\newcommand{\uKLI}{^{\sss KLI}}

\begin{document}
\title{Density-functional approach to the band gaps of finite and periodic two-dimensional systems}
\author{Alberto Guandalini}
\email{alberto.guandalini@unimore.it}
\affiliation{Dipartimento di Scienze Fisiche, Informatiche e Matematiche, Universit{\`a} di Modena e Reggio Emilia, Via Campi 213A, I-41125 Modena, Italy}
\affiliation{CNR -- Istituto Nanoscienze, Via Campi 213A, I-41125 Modena, Italy}
\author{Alice Ruini}
\affiliation{Dipartimento di Scienze Fisiche, Informatiche e Matematiche, Universit{\`a} di Modena e Reggio Emilia, Via Campi 213A, I-41125 Modena, Italy}
\affiliation{CNR -- Istituto Nanoscienze, Via Campi 213A, I-41125 Modena, Italy}
\author{Esa R\"as\"anen}
\affiliation{Computational Physics Laboratory, Tampere University, P.O. Box 692, FI-33101 Tampere, Finland}
\author{Carlo A. Rozzi}
\affiliation{CNR -- Istituto Nanoscienze, Via Campi 213A, I-41125 Modena, Italy}
\author{Stefano Pittalis}
\email{stefano.pittalis@nano.cnr.it}
\affiliation{CNR -- Istituto Nanoscienze, Via Campi 213A, I-41125 Modena, Italy}

\begin{abstract}
We present an approach based on density-functional theory for the calculation of fundamental gaps of both finite and  periodic  two-dimensional (2D) electronic systems. The computational cost of our approach is comparable to that of total energy calculations performed via standard semi-local forms.
We achieve this by  replacing the 2D  local density approximation with a more sophisticated -- yet computationally simple -- orbital-dependent modeling of the exchange potential within the procedure by Guandalini {\it et al.} [Phys. Rev. B {\bf 99}, 125140 (2019)]. We showcase promising results for  semiconductor 2D quantum dots and artificial graphene systems, where the band structure can be tuned through, e.g., Kekul\'e distortion. 
\end{abstract}

\maketitle

\section{Introduction}
During the past decades, the ability to produce two-dimensional (2D) electron gas has led to the discovery of new physical phenomena and applications,  including integer and fractional quantum Hall effect, semiconductor nanodevices such as quantum dots, and a variety of topological quantum systems~\cite{martin2020}. On the other hand, the discoveries of atomic 2D materials, such as graphene, have inspired band-structure engineering in a variety of 2D systems~\cite{Park2009, Gibertini2009, Singha1176,Rasanen2012,Gomes2012, Paavilainen2016,Tarruell2012,Polini2013,slot2017,Khajetoorians2019}. 
In addition, the recently discovered properties of twisted bilayer graphene~\cite{bistritzer2011,cao2018unconventional,cao2018b} have led to next phase of so-called designer materials~\cite{kim2017,chen2019,tartakovskii2019,yankowitz2019,lu2019,stepanov2020,ReserbatPlantey2021}.

Density-functional theory (DFT) can provide a practical yet sufficiently accurate approach to compute the electronic structure of 3D materials and quantum confined low-dimensional systems~\cite{Note1}.
The reliability of the results depends on the choice to approximate the exchange-correlation (xc) energy functional~\cite{Perdew2003,SS05,MHG17}, which accounts for the many-body effects beyond the (classical) Hartree interaction.
To examine 2D systems, the standard reference to the 3D electron gas used to derived approximate semi-local functional forms for atoms, molecules, and solids must be replaced by the 2D electron gas. Although numerous forms have been derived by now for 2D systems,~\cite{Attaccalite2002, PRHG07,PhysRevB.78.195322,pittalis2009density,RPPG09,pittalis2009correlation,PhysRevA.80.032515,PhysRevB.80.165112,RASANEN20101232,Pittalis2010a,PhysRevB.81.195103,PhysRevB.82.195124,Pittalis2010,Vilhena2014,SP2017,ASP2018,ASP2018B,JPP2018,PP2019} the computation of band gaps sets additional, non-trivial, challenges.

From the experience on 3D materials, it is known that standard semi-local DFT approximations (DFAs)  often underestimate the actual band gaps, because they miss a crucial discontinuous shift in the xc potential. Accurate band gaps can be obtained from advanced many-body approximations such as GW calculations~\cite{GSS88,AG98,ORR02,GMR06}, or by using orbital-dependent DFT functionals, e.g., hybrids~\cite{HSE03,EHSE,SEKB10,KSRB12,Franchini14,Periodic17}. However, these approaches are computationally demanding. Further progress can  be made by upgrading semi-local DFAs to consistent ensemble-DFT (EDFT) forms~\cite{KK13,kk14,AG15,GMB16}, by switching to a recently introduced ``N-centered'' EDFT~\cite{SF18,SF20,HWF21}, or via non-variational models of the xc potential which can balance simplicity and accuracy~\cite{Gritsenko1995,TB09,KOER10,Kuisma2010,H13,AK13,COM14,C1EE02717D,STATOVEROV14,EJB17,TSB18,LargeScaleBenchmark19,RMB20}. The latter option is the thread that we follow here.

In this work, we extend the approach presented in Ref.~\onlinecite{Guandalini2019} -- which focused on finite systems -- to compute the fundamental gaps of both finite {\it and} periodic 2D systems. We keep the computational cost restricted at about the level  of a ground-state calculation performed by means of a standard semi-local approximation, then followed by an additional iteration at almost no extra cost.
As a key step beyond what done in Ref.~\onlinecite{Guandalini2019}, we ``transfer'' the GLLB model  potential by Gritsenko {\it et al.}~\cite{Gritsenko1995} and its modifications by Kuisma {\it et al.}~\cite{Kuisma2010} from the domain of 3D atomistic materials to 2D semiconductor devices.

The details of the extended modeling and the employed procedure, including a brief introduction to the relevant aspects of DFT, are described in 
Sec.~\ref{SecThGLLB}. In Sec. \ref{SecApplGLLB} we demonstrate the performance of the approach in two relevant sets of applications. They comprise finite 2D quantum dots and two types of periodic 2D systems, i.e., artificial graphene and its Kekulé variant.  Our conclusions are summarized in Sec.~\ref{SecConc}.

\section{Theory}\label{SecThGLLB}
In the Kohn-Sham (KS) DFT approach, the total energy of a 2D system with $N$ electrons is expressed as a functional of the particle density, $n$,
as follows
\begin{equation}\label{2DGLLBEqEtot}
E[n] = T_{\mathrm{KS}}[n]+\int d^2 r\; v_{0}(\vr)n(\vr)+E_{\mathrm{H}}[n]+E_{\mathrm{xc}}[n],
\end{equation}
where $T_{\mathrm{KS}}[n]$ is the KS kinetic energy,
$v_{0}$ is the external potential of the system (explicit examples are given in Section \ref{SecApplGLLB}),
$E_{\rm H}$ is the Hartree energy, and $E_{\rm xc}$ is the xc energy of the 2D system. The KS equation reads
\begin{multline}\label{EqKSEqs}
\left[-\frac{\nabla^2}{2}+v_{0}(\vr)+v_{\mathrm{H}}[n](\vr)+v_{\mathrm{xc}}[n](\vr)\right]\varphi_{j}(\vr) \\
=\epsilon_{j}\varphi_{j}(\vr),
\end{multline}
where
\begin{equation}
v_{\mathrm{H}}[n](\vr) = \int d^2r' \frac{n(\vr')}{|\vr-\vr'|}
\end{equation}
is the Hartree potential and 
\begin{equation}\label{EqGLLBvxcdef}
v_{\mathrm{xc}}[n](\vr) = \frac{\delta E_{\mathrm{xc}}[n]}{\delta n(\vr)}
\end{equation}
is the xc potential. The {\it exact} $E_{\rm xc}$ and $v_{\mathrm{xc}}$ subsume the effects of the electron-electron interaction including those beyond simple mean-field modeling.
The electronic density is  obtained from the KS orbitals as
\begin{equation}
n(\vr) = \sum\limits_{j=1}^{occ.}|\varphi_{j}(\vr)|^2,
\end{equation}
where the index $j$ of the occupied single-particle KS orbitals 
$\varphi_j$ includes spin.

An approximation for $v_{\mathrm{xc}}[n]$ can be obtained from the functional derivative of an approximate $E_{\mathrm{xc}}$ according to Eq.~(\ref{EqGLLBvxcdef}). 
In this work, instead, we directly model $v_{\mathrm{xc}}$  for practical purposes. 
Preserving the variational character of the exact $v_{\mathrm{xc}}$ is not central in this work because 
we are after energies differences at {\it prescribed} system geometries. These energy differences can be determined via KS eigenvalues directly as specified in Sec. \ref{Disc}.

\subsection{Modeling the xc potential}\label{SecXCv}

We begin with considering the exchange component $v_{\mathrm{x}}$ of the xc potential $v_{\mathrm{xc}}$ (notation $[n]$ omitted for brevity below). The Krieger-Li-Iafrate (KLI) potential\cite{Krieger1992} (usually) constitutes a useful approximation for exact $v_{\mathrm{x}}$, both for finite and extended systems~\cite{kk08,TBBB16}. 
In 2D, the KLI expression is given by
\begin{equation}\label{EqGLLBKLIEq}
v_{\mathrm{x}}^{\mathrm{KLI}}(\vr) = v_\mathrm{Sl}(\vr) + 
\sum\limits_{i=1}^{occ.} w_{i}\frac{|\varphi_{i}(\vr)|^2}{n(\vr)} 
\end{equation}
where 
\begin{equation}\label{SL}
v_\mathrm{Sl}(\vr) = \frac{1}{ n(\vr)}\sum\limits_{i=1}^{occ.}  |\varphi_{i}(\vr)|^2  u_{\mathrm{x}i}(\vr) 
\end{equation}
with
\begin{equation}
u_{\mathrm{x}i}(\vr) = -\sum\limits_{j=1}^{occ.}\frac{\varphi_{j}(\vr)}{\varphi^*_{i}(\vr)}\int d^2r' \frac{\varphi^*_{i}(\vr')\varphi_{j}(\vr')}{|\vr-\vr'|}\;,
\end{equation}
is the Slater (Sl) potential. The term $w_i$ in Eq.~\ref{EqGLLBKLIEq} is written as
\begin{equation}\label{EqGLLBwi}
w_{i} = \int d^2 r \left[v_{\mathrm{x}}^{\mathrm{KLI}}[n](\vr)-u_{\mathrm{x}i}[n](\vr)  \right] |\varphi_{i}(\vr)|^2\;.
\end{equation}

According to the (3D) GLLB approach~\cite{Gritsenko1995}, the key features of the KLI potential can be captured by a computationally simple model potential. In 2D, we first consider the second term in Eq.~(\ref{EqGLLBKLIEq}). This term is crucial as it exhibits a non-vanishing discontinuity at an integer electron number (more below). Therefore, we replace $w_{i}$ in Eq.~(\ref{EqGLLBwi}) by
\begin{equation}\label{w2DGLLB}
w_{i} \rightarrow w\uGLLB_{i} \equiv K_{\mathrm{x}}\uTW\sqrt{\mu-\epsilon_{i}};
\end{equation}
where $\mu$ is the chemical potential and $K_{\mathrm{x}}\uTW = \sqrt{2}/\pi \approx 0.4502$ is a constant determined to exactly reproduce the case of the homogeneous 2D electron gas. In particular, 2D local-density approximation (2DLDA) corresponds to
$v_{\mathrm{x}}\uLDA = -\frac{2}{\pi}\sqrt{2\pi n}$ and
$ v_{\mathrm{Sl}}\uLDA = 2\epsilon_{\mathrm{x}}\uLDA=\frac{4}{3}v_{\mathrm{x}}\uLDA
$ (Ref.~\cite{Rajagopal1977}).

Next, analogously to the original GLLB model potential, we approximate the Slater potential with Becke's B88 exchange functional~\cite{Becke1988}. However, here we use the 2D version of the B88 (2DB88) derived in Ref.~\cite{Vilhena2014}. Thus, we set
\begin{equation}\label{2DB88}
v_{\rm Sl} \rightarrow v_\mathrm{Sl}\uGLLB \equiv  2 \epsilon_{\mathrm{x}}\uTWBK\;.
\end{equation}
This approximation correctly captures the long-range limit $v_{\rm Sl}(r >> 1) \sim -1/r$. This feature is of particular importance in finite systems in order to correctly describe the tail of the xc potential. 

In periodic systems,  the aforementioned long-range behavior is still pertinent for well separated centers. Yet, it may overall be expected to play a less prominent role than for finite systems. Here, we follow an adaptation of the GLLB to atomistic 3D periodic systems which is known under the acronym GLLB-SC, where "SC" stands for solids and correlation~\cite{Kuisma2010}. The GLLB-SC involves a PBE-like approximation (a generalized-gradient approximation) for solids~\cite{Perdew2008}; i.e., PBEsol. The difference between PBEsol and  the regular PBE is essentially in the way the gradient corrections are weighted. 

However, PBEsol and PBE approximations in 2D are -- to the best of our knowledge -- not available. Thus, partially sacrifying higher accuracy that may be achieved, we shall ignore gradient corrections in $\epsilon_{\mathrm{x}}$ for the periodic cases.  Furthermore, we shall also 
use the 2DLDA for the correlation potential~\cite{Rajagopal1977,Tanatar1989,Attaccalite2002} in all the cases.

In summary, we propose
\begin{align} \label{EqGLLBtotPotXC}
 v_{\mathrm{xc}}\uGLLBpSCp &= 2\epsilon_{\mathrm{x}}\uTWBKpLDAp + K_{\mathrm{x}}\uTW \sum\limits_{i=1}^{occ.}  \sqrt{\mu-\epsilon_{i}}\frac{|\varphi_{i}|^2}{n} \nonumber \\
& + v_{\mathrm{c}}\uLDA\;,
\end{align}
where grouping of the terms and notation intend to emphasize that the difference between these versions of the 2DGLLB and 2DGLLB-SC is  confined within the way $\epsilon_{\mathrm{x}}$ is approximated. Although, as explained above, the present version of the 2DGLLB-SC is not yet fully optimal as comparing as to the 3D analog~\cite{Kuisma2010}, it may be further improved.

\subsection{Discontinuity of the model potential}\label{Disc}
In EDFT~\cite{PPLB82,PhysRevLett.51.1884,SS85}, the fundamental gap can be expressed as the sum of two contributions:
\begin{equation}\label{EqGLLBFundGap}
G_{\Delta} = \Delta_{\mathrm{KS}} + \Delta_{\mathrm{xc}},
\end{equation}
where $\Delta_{\mathrm{KS}}$
is the KS gap and
\begin{equation}\label{Deltaxc}
\Delta_{\mathrm{xc}} = \lim\limits_{\delta N \to 0^+} \left\lbrace \left.v_{\mathrm{xc}}(\vr)\right|_{N+\delta N}-\left.v_{\mathrm{xc}}(\vr)\right|_{N-\delta N} \right\rbrace
\end{equation}
is referred to as the discontinuity of the xc potential. 

Using Eq.~(\ref{EqGLLBtotPotXC}) in Eq.~(\ref{Deltaxc}), we obtain
\begin{multline}\label{EqGLLBDerDiscr}
\Delta_{\mathrm{xc}}\uGLLBpSCp(\vr) = 
K_{\mathrm{x}}\uTW  \\
\times \sum\limits_{i=1}^{occ.}\ \left( \sqrt{\epsilon_{\mathrm{L}}-\epsilon_{i}}-\sqrt{\epsilon_{\mathrm{H}}-\epsilon_{i}} \right)  \frac{|\varphi_{i}(\vr)|^2}{n(\vr)},
\end{multline}
where $\epsilon_{\mathrm{L}}$ ($\epsilon_{\mathrm{H}}$) stands for the lowest (highest) unoccupied (occupied) single particle level; i.e., the bottom (top) of the conduction (valence) band. Note that $\Delta_{\mathrm{KS}} = \epsilon_{\mathrm{L}}-\epsilon_{\mathrm{H}}$.
Note that the $\Delta_{\mathrm{xc}}\uGLLBpSCp$ in Eq.~(\ref{EqGLLBDerDiscr})
originates solely from the term of  Eq.~(\ref{EqGLLBtotPotXC}) which  depends explicitly on the orbitals via
$\sqrt{\mu-\epsilon_{i}}\frac{|\varphi_{i}|^2}{n}$.
Yet, all the terms of the xc-potential in Eq.~(\ref{EqGLLBtotPotXC}) can affect any finite value of $\Delta_{\mathrm{xc}}$.

It is noteworthy that the expression in Eq.~(\ref{EqGLLBDerDiscr}) is
position-dependent, while the exact discontinuity is a constant. 
Therefore, first-order perturbation theory was employed by
Kuisma {\it et al.}~\cite{Kuisma2010} in the modification of the GLLB result for 
band gaps. Following the same approach in 2D,  we get
\begin{equation}\label{EqGLLBDDKuisma}
\bar{\Delta}\uGLLBpSCp_{\mathrm{xc}} \equiv \mel{\varphi_{\mathrm{L}}}{\Delta_{\mathrm{xc}}\uGLLBpSCp(\vr)}{\varphi_{\mathrm{L}}}.
\end{equation}

However, according to our tests (not reported here) Eq.~(\ref{EqGLLBDDKuisma})
fails to reproduce accurate fundamental gaps for single quantum dots.
To remedy this issue, we evaluate $\Delta_{\mathrm{xc}}$ by following the procedure detailed in Refs.~\onlinecite{Guandalini2019,EJB17}. It may be useful to summarize below the key steps. \\

First, at the level of the KLI approximation for exact-exchange only, we have the {\it exact} expression~\cite{Guandalini2019}
\begin{equation}\label{KLIDD}
\Delta_{\mathrm{x}}\uKLI = \tilde{\epsilon}_{\mathrm{H},N+1}\uKLI-\epsilon\uKLI_{\mathrm{L},N}.
\end{equation}
Here, the superscript ($ \sim $) indicates that the frozen orbital approximation must be employed in computing
the potential that generates $\tilde{\epsilon}_{\mathrm{H},N+1}$. 
Thus, the evaluation of $\Delta_{\mathrm{x}}\uKLI$ only requires a one-shot iteration for a system with $N+1$ electrons, so that the converged
$N$-electron ground state is used as the initial state for this one-shot computation. 

Next, we may invoke the LDA in Eq.~(\ref{KLIDD}) as follows
\begin{equation}\label{LDADD}
\Delta_{\mathrm{xc}}\uLDA = \tilde{\epsilon}_{\mathrm{H},N+1}\uLDA-\epsilon\uLDA_{\mathrm{L},N}.
\end{equation}
We point out that in this step we also {\it assume} that the correlation may be treated in a  straightforward fashion as above. This yields {\it non-}vanishing contributions for finite systems and thus rather accurate fundamental gaps~\cite{Guandalini2019}.

For the 2DGLLB(-SC) model potential, we readily get
\begin{equation}\label{EqGLLBOurDD} 
\Delta_{\mathrm{xc}}\uGLLBpSCp = \tilde{\epsilon}_{\mathrm{H},N+1}\uGLLB-\epsilon\uGLLBpSCp_{\mathrm{L},N}.
\end{equation}
Equations~(\ref{KLIDD}), (\ref{LDADD}), and (\ref{EqGLLBOurDD}) involve eigenvalues, which are obtained by solving the 
KS equations for the corresponding approximate xc-potential.
Equation~(\ref{EqGLLBOurDD}) differs from Eq. (\ref{EqGLLBDDKuisma}) by
\begin{multline}\label{diff}
\Delta_{\mathrm{xc}}\uGLLBpSCp-\bar{\Delta}\uGLLBpSCp_{\mathrm{xc}} =\\
\mel{\varphi_{\mathrm{L}}}{v_{\mathrm{H}}[n+n_{\mathrm{L}}]-v_{\mathrm{H}}[n]}{\varphi_{\mathrm{L}}}
+ \bra{\varphi_{\mathrm{L}}} K_{\mathrm{x}}\uTW \\
\times\sum\limits_{i=1}^{occ.}\sqrt{\epsilon_{\mathrm{L}}-\epsilon_{i}}  \left[\frac{|\varphi_{i}(\vr)|^2}{n(\vr)+n_{\mathrm{L}}(\vr)}-\frac{|\varphi_{i}(\vr)|^2}{n(\vr)}\right]\ket{\varphi_{\mathrm{L}}} \ .
\end{multline}
Yet, for extended periodic systems, Eq.~(\ref{EqGLLBOurDD}) can reduce to  Eq. (\ref{EqGLLBDDKuisma}). The proof given for regular 3D cases by Baerends~\cite{EJB17} applies with minor modifications to analogous 2D cases as well. The key assumption is that an electron added to an extended system will spread over the entire  structure. Then in Eq.~(\ref{diff}),  $n+n_L \rightarrow n$ in the macroscopic limit.
In the same limit, it should be noted that Eq.~(\ref{LDADD}) yields $\Delta_{\mathrm{xc}}\uLDA \rightarrow 0$. This problem can be avoided by using the proposed 2DGLLB-SC model.

\section{Applications}\label{SecApplGLLB}
We have implemented the 2DGLLB(-SC) potential in a local version of the OCTOPUS software package~\cite{octopus15,octopus06,octopus03}. The OCTOPUS code
solves the KS equations on a regular grid with Dirichlet or periodic boundary conditions without being bounded to any particular choice of the basis set. All the fundamental gaps shown in this work are converged to the fourth significant digit.

\subsection{Finite systems: Semiconductor quantum dots}

\subsubsection{Model and parameters}

As our first application of the proposed approach we examine the fundamental gaps of single quantum dots (QDs) containing $N$ interacting electrons. We consider the conventional model for semiconductor QDs~\cite{RM02} by using the effective mass approximation with the parameters for GaAs, i.e., 
$m^*=0.067m_e$ and $\epsilon=12.4\epsilon_0$, and a harmonic confining potential with an elliptic deformation. In effective atomic units (eff. a.u.) used here and throughout we have
\begin{equation}\label{EqGLLBQD}
v_0(\vr) = \frac{1}{2}\omega^2(x^2+\alpha^2y^2),
\end{equation}
where $\omega$ determines the strength of the confinement and $\alpha$ defines an elliptical deformation. If $\alpha \neq 1$, the degeneracy 
of the single-particle levels is removed.\\
The simulation box containing the real-space domain is a circular cavity with a radius 
of $R=K/\sqrt{\omega}$, where $K=5.0$ is used for $N=2,4,5$ and $K=\lbrace 6.0, 6.5, 7.0, 7.5, 8.0, 8.5\rbrace$ is used for $N=\lbrace 6,12,20,30,42,56 \rbrace$, respectively. 
The grid spacing is $g=0.1/\sqrt{\omega}$.

\subsubsection{Results}

\begin{figure}
\begin{center}
\includegraphics[width=1.00 \linewidth,=1]{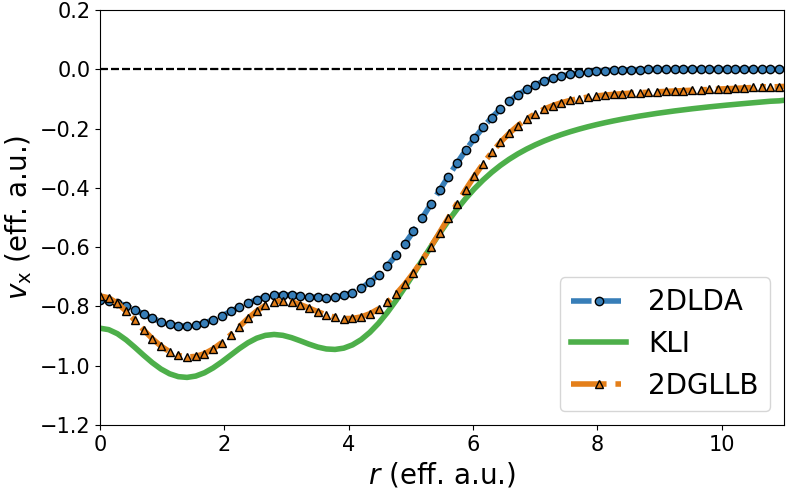}
\caption{Non-self-consistent exchange potential, $v_{\rm x}$, of a quantum dot with $N=20$, $\alpha = 1$ and $\omega = 0.5$ computed within different approximations. 
}~\label{FigPotUnif}
\end{center}
\vskip -0.5cm
\end{figure}

Figure~\ref{FigPotUnif}
shows the 2DGLLB, KLI, and 2DLDA exchange potentials, $v_x$, as a function of the QD radius $r$ computed for a QD with
 $\alpha = 1$, $N=20$, and $\omega = 0.5$.
The 2DGLLB and 2DLDA potentials are evaluated non-self-consistently for the converged KLI ground state. We find that both the  2DGLLB and 2DLDA overestimate the KLI potential, but the 2DGLLB is considerably closer to the KLI than the 2DLDA. The condition $v_{\mathrm{x}}\uGLLB \ge v_{\mathrm{x}}\uKLI$ is due to the fact that $w_i\uGLLB \ge w_i$ in
Eq.~\ref{w2DGLLB}. We also note that the shell structure of the QD is better reflected in the 2DGLLB potential than in the 2DLDA potential.

\begin{figure}
\includegraphics[width=1.0 \linewidth,=1]{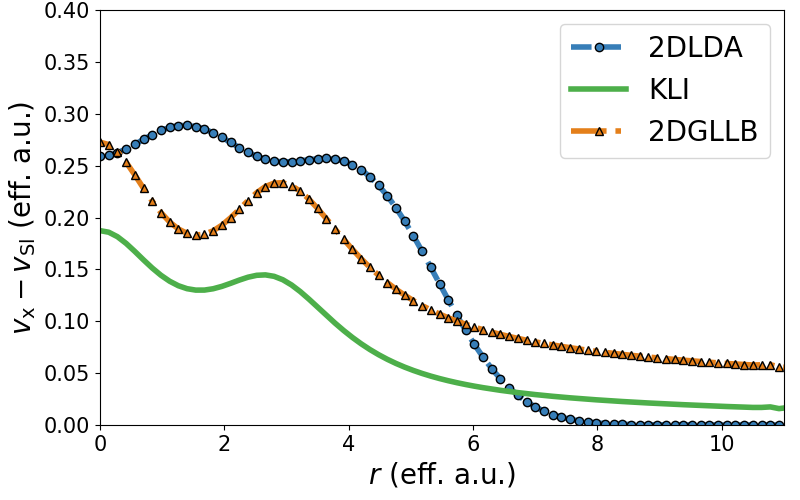}
\caption{Same as Fig. \ref{FigPotUnif} but for the non-Slater part of the exchange potential; i.e., $v_{\mathrm{x}} - v_{\rm Sl}$.}
\label{FigPotSep}
\vskip -0.5cm
\end{figure}

Figure~\ref{FigPotSep} shows the difference $v_{\mathrm{x}} - v_{\mathrm{Sl}}$ for the same case as in Fig.~\ref{FigPotUnif}. The 2DLDA is both quantitatively and qualitatively different from the KLI. In particular,
2DLDA potential cannot capture the correct shell structure of the QD. In spite of an almost constant shift, the 2DGLLB result resembles the KLI, whereas the 2DLDA has a flatter profile.

\begin{figure}
\begin{center}
\includegraphics[scale=0.43]{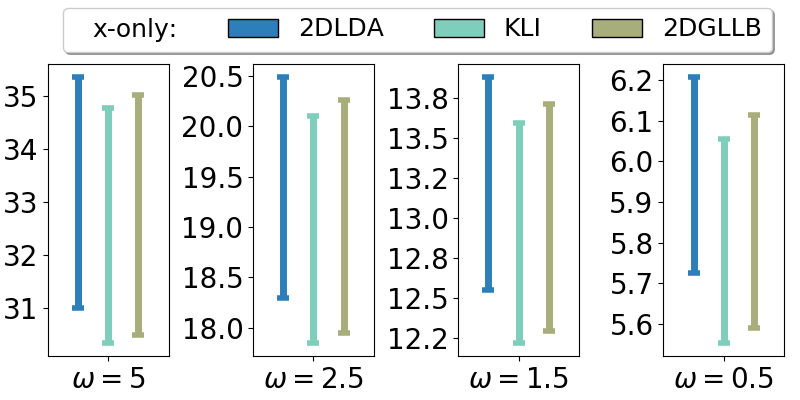}
\caption{Ionization potential (bottom bar value) and electron affinity (top bar value) of elliptic quantum dots with $N=12$ and $\alpha = 1.05$ for varying confinement strengths $\omega$ computed within different approximations. 
The upper (lower) values of the bars correspond to electron affinities (ionization potentials), and the length of the bars corresponds to the  fundamental gaps. Correlation contributions are ignored for all cases.}
\label{FigGLLBQDsO_IPA}
\end{center}
\vskip -0.5cm
\end{figure}

\begin{figure*}
\begin{center}
\includegraphics[scale=0.40]{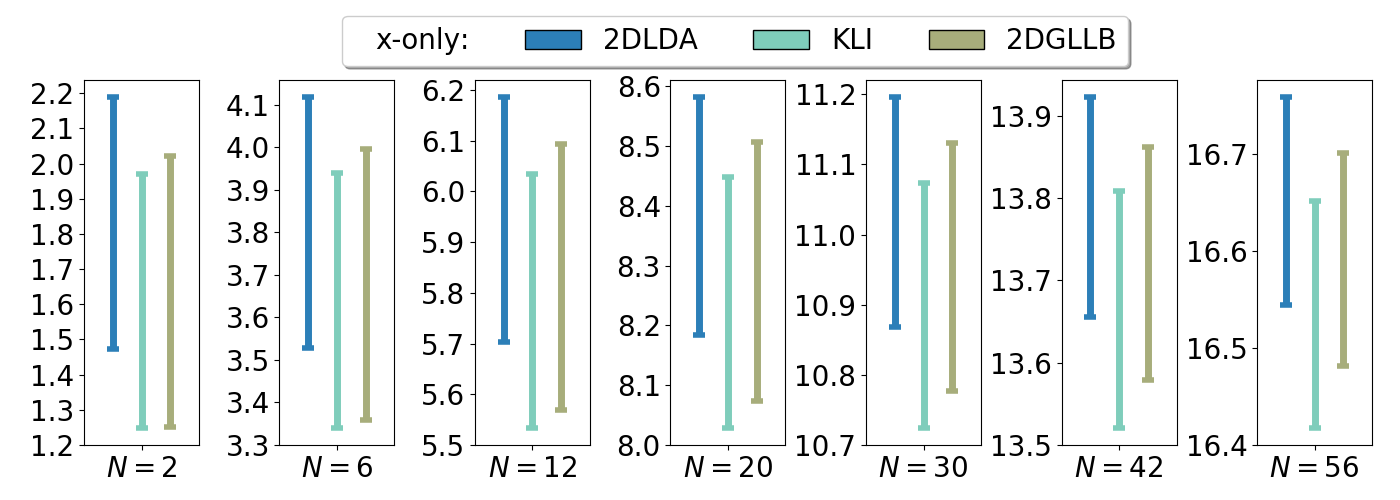}
\caption{Same as  Fig.~\ref{FigGLLBQDsO_IPA} but for a fixed value of the confinement strength $\omega = 0.5$ and varying number of electrons $N$.}\label{FigGLLBQDsN_IPA}
\end{center}
\vskip -0.5cm
\end{figure*}

Next, let us recall that the ionization potential  and the electron affinity are give by $I \equiv \epsilon_H$ and $A \equiv \epsilon_L + \Delta_{\mathrm{xc}}$, respectively. Here, $\epsilon_H$ refers to energy of the highest occupied KS state and $\epsilon_L$ to that of the lowest unoccupied KS state. According to Eq.~\eqref{EqGLLBFundGap}, we thus have $G_{\Delta} = A - I$ for the fundamental gap. As explained in Sec.~(\ref{Disc}),  $\Delta_{\mathrm{xc}}$ can be evaluated according to Eq.~(\ref{LDADD}) in the 2DLDA case, and Eq.~(\ref{EqGLLBOurDD}) in the 2DGLLB(-SC) case.

Figures~\ref{FigGLLBQDsO_IPA} and \ref{FigGLLBQDsN_IPA} visualize all the aforementioned quantities for QDs with varying $\omega$ and $N$, respectively, computed at the level of various exchange-only (x-only) approximations.
Using, again, the KLI results as a reference, we note that the 2DLDA overestimates both the 
ionization potential and electron affinity for all the QDs. This is due to the inaccurate KS eigenvalues $\epsilon_H$ and $\epsilon_L$ resulting from the erroneous long-range behavior of the 2DLDA x-potential. 
We also note that the 2DLDA tends to underestimate the fundamental gaps,
even though the errors in $A$ and $I$ are at least partly canceled out in $G_\Delta$. 
In 2DGLLB instead the ionization potentials and electron affinities are improved individually. This can be ascribed to the improved description of Slater potential (especially of its tail) obtained via the 2DB88 approximation. The 2DGLLB slightly overestimates (underestimates) the fundamental gaps at lower $N$ (at higher $N$). 
Overall, the accuracy of the 2DGLLB for the fundamental gaps is comparable to that of the 2DLDA.  

In detail, the mean absolute error for the ionization potentials is $0.06$  ($0.25$) in the  2DGLLB (2DLDA) case. For the electron affinities the error is $0.09$ ($0.22$) in the 2DGLLB (2DLDA) case. Finally, for the the fundamental gaps the error is $0.03$ ($0.03$) in the 2DGLLB (2DLDA) case.
Furthermore -- in line with the expectation that the 2DGLLB-SC  potential is less adapted than the 2DGLLB potential to deal with finite systems -- the 2DGLLB-SC  mean absolute errors are $0.16$,  $0.18$, and $0.04$, respectively. We will consider the 2DGLLB-SC case more in detail for 2D periodic systems in the next section.

\begin{figure}
\begin{center}
\includegraphics[width=1.0 \linewidth,=1]{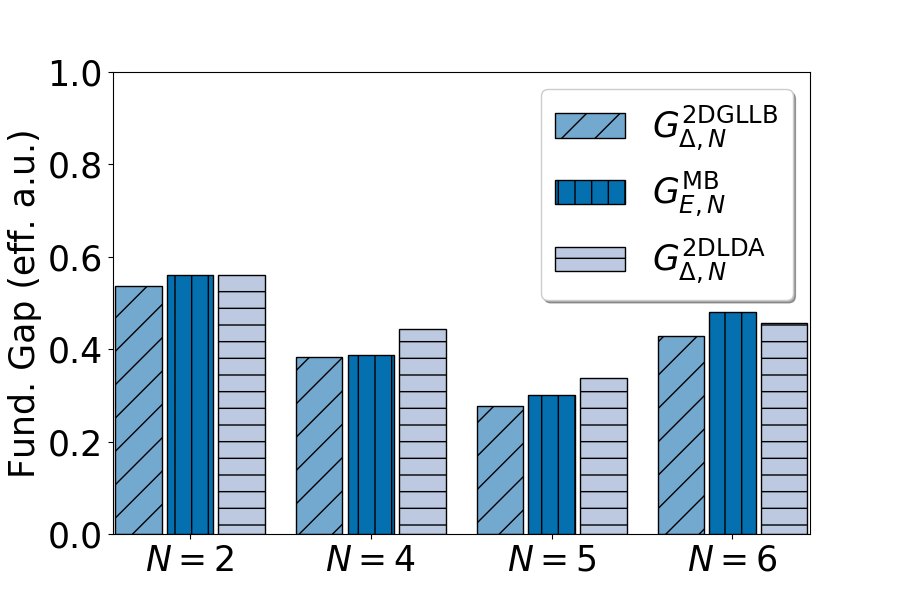}
\caption{Fundamental gaps for quantum dots with $\alpha = 1$, $\omega = 0.35$ and varying number of electrons $N$ with correlation contributions included. $G_{E,N}^{\mathrm{MB}}$ is the full configuration interaction value from  Ref.~\onlinecite{CBKR07}; $G_{E,N}^{\mathrm{2DLDA}}$ is obtained with 2DLDA as explained in Ref. \onlinecite{Guandalini2019}; $G_{E,N}\uGLLB$ is obtained with the 2DGLLB approximation according to Eq. \eqref{EqGLLBOurDD}.}\label{FigGLLBQDsN_corr}
\end{center}
\vskip -0.5cm
\end{figure}

Finally, we examine cases for quantum dots  {\it including} correlations.  
Here we consider a set of 
parabolic QDs ($\alpha = 1$) with $\omega = 0.35$ and $N = 2\ldots 6$ for which numerically exact configuration interaction results are available~\cite{CBKR07}. Our results are shown in Fig.~\ref{FigGLLBQDsN_corr}. Overall, both 2DGLLB and 2DLDA yield 
accurate results in all considered cases. In particular, the 2DGLLB slightly  underestimates the gap for all cases, while the 2DLDA underestimates the gap for closed-shell systems ($N=2,6$) and overestimates the gap in the open-shell systems ($N=4,5$). 
 Both $A$ and $I$ are systematically more accurate in 2DGLLB than in 2DLDA.
The mean absolute error for the fundamental gaps is $0.025$ and $0.029$ for the 2DGLLB and 2DLDA case, respectively.

\subsection{Periodic systems: Artificial graphene}

\subsubsection{Model and parameters}

\begin{figure}
\begin{center}
\includegraphics[width=1.0 \linewidth,=1]{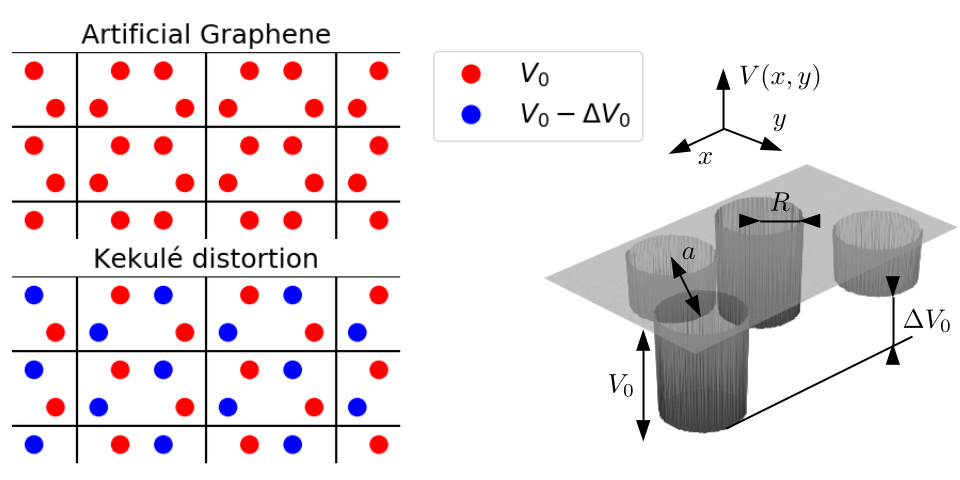}
\caption{Left: Lattice of artificial graphene (upper panel) and that with the Kekulé distortion (lower panel). Right: Rectangular unit cell containing four pillar-like quantum dots with depth $\Delta V_0$. The used parameter values are $a = 150$ nm, $V_0 = 0.60$ meV, $R = 52.5$ nm, and $\Delta V_0 = 0\ldots 0.40$ meV.}\label{FigPeriodicSystem}
\end{center}
\vskip -0.5cm
\end{figure}

To test the proposed approached in periodic 2D systems, we consider artificial graphene~\cite{Park2009,Gibertini2009,Singha1176,Rasanen2012,Gomes2012,Paavilainen2016,Tarruell2012,Polini2013,Du2018}, (AG) and its Kekul\'e distortion leading to a band-gap opening~\cite{Gomes2012,Paavilainen2016}. In particular, we focus on AG realized
in nanopatterned 2D electron gas in GaAs heterostructures~\cite{Park2009,Gibertini2009,Singha1176,Rasanen2012,Du2018} that can be modeled in real space with a hegagonal array of QDs~\cite{Rasanen2012,Kylanpa2016}. This system shows Dirac cones similar to conventional graphene, and the band structure can be tuned at will through, e.g., Kekul\'e distortion.

In Fig.~\ref{FigPeriodicSystem}, we show a schematic figure of the AG (left) and its Kekul\'e distortion, together with the rectangular unit cells for both cases. The calculations were performed in a non-primitive cell due to numerical restrictions, and thus the bands had to be unfolded as described in Refs.~\onlinecite{Rasanen2012} and \onlinecite{Kylanpa2016}. The lattice constant is set to $a = 150$ nm, and the QDs are modeled by a cylindrical hard-wall potential with radius $R=52.5$ nm. The grid spacing is set to $\approx 2.45$ nm. The potential depth of each QD is $V_0 = 0.60$ meV, and in the Kekul\'e case two of the four QDs in the unit cell are lowered by up to $\Delta V_0=0.4$ meV. We use the same GaAs parameters as in the previous section ($m^*=0.067m_0$ and $\epsilon = 12.4 \epsilon_0$). We set four electrons in a unit cell, so that each QD is occupied by one electron. Such a low-density system is experimentally viable~\cite{Hirjibehedin2002,Hirjibehedin2005,Garcia2005}. 

The irreducible Brillouin zone is sampled with a $12 \times 12$ regular grid according to  a modified Monkhorst-Pack scheme~\cite{Pack1977}.
In order to compute the band structure, we sample $100$ equally spaced $k$ points along the path $\Gamma$-$M$-$K$-$\Gamma$. Here, we will show energies in meV and lengths in nm in accordance with the previous works~\cite{Gibertini2009,Rasanen2012,Kylanpa2016}.

\subsubsection{Results}

\begin{figure*}[t]
\includegraphics[width=\textwidth]{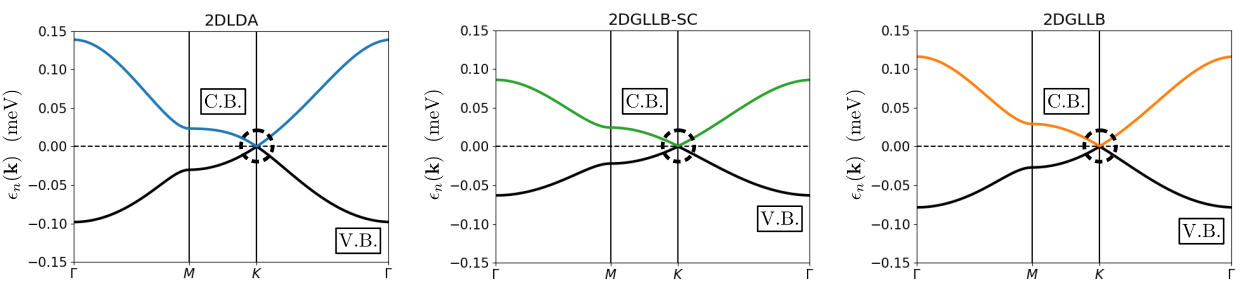}
\includegraphics[width=\textwidth]{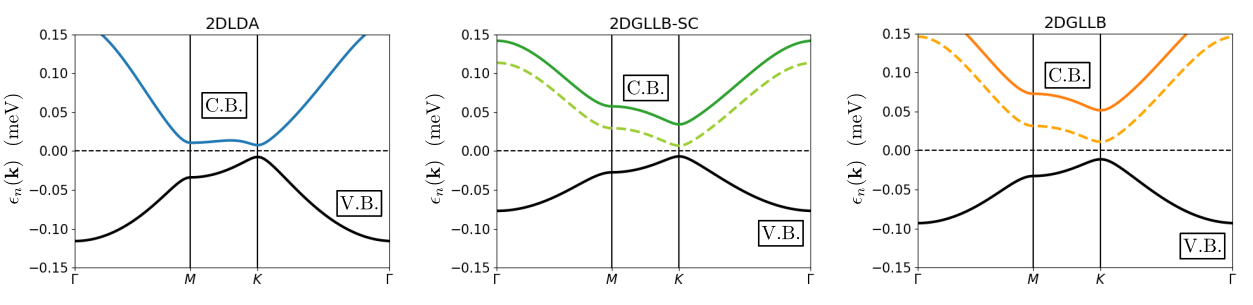}
\caption{Energy bands of artificial graphene (upper row) and that with the Kekulé distortion of $\Delta V_0 = 0.30$ meV (lower row) along the symmetry path $\Gamma$-$M$-$K$-$\Gamma$ obtained with the 2DLDA (left column),
 2DGLLB-SC (center column), and 2DGLLB (right column). The valence and conduction bands are labeled as V.B. and C.B., respectively. The Fermi energy (dashed line) is set to zero. The solid lines show the actual conduction bands, and the dashed lines show the bare Kohn-Sham bands.}\label{FigArtGraphBands}
\vskip -0.5cm
\end{figure*}

In the upper panel of Fig.~\ref{FigArtGraphBands}, we show the energy bands of the AG computed with 2DLDA, 2DGLLB-SC, and 2DGLLB, respectively. As expected, all the cases show Dirac cones with linear dispersion relation at the $K$ point. Since AG is a periodic system, $\Delta_{\mathrm{xc}}^{\mathrm{LDA}} = 0$. Because AG is a semimetal (with zero band gap), the top of the valence band and the bottom of the conduction band have the same energy, i.e., $\epsilon_{\mathrm{L}}=\epsilon_{\mathrm{H}}$. Thus, also $\Delta_{\mathrm{xc}}\uGLLBpSCp = 0$. The 2DGLLB(-SC) results show slightly smaller band dispersion than the 2DLDA counterpart.

In the lower panel of Fig.~\ref{FigArtGraphBands}, we show the energy bands of AG with Kekul\'e distortion ($\Delta V_0 = 0.30$ meV). As expected, the Kekul\'e distortion opens a gap at the $K$ point. The valence bands are qualitatively similar in 2DGLLB(-SC) and 2DLDA.
However, as in the case of regular AG, the 2DGLLB(-SC) bands are less dispersive than in the LDA case. The conduction bands shows more differences. In particular, $\Delta_{\mathrm{xc}}^{\mathrm{2DLDA}} = 0$, while $\Delta_{\mathrm{xc}}\uGLLBpSCp \ne 0$. This fact significantly affects the position of the conduction bands with respect to the bare KS bands.

\begin{figure}
\begin{center}
\includegraphics[width=1.00 \linewidth,=1]{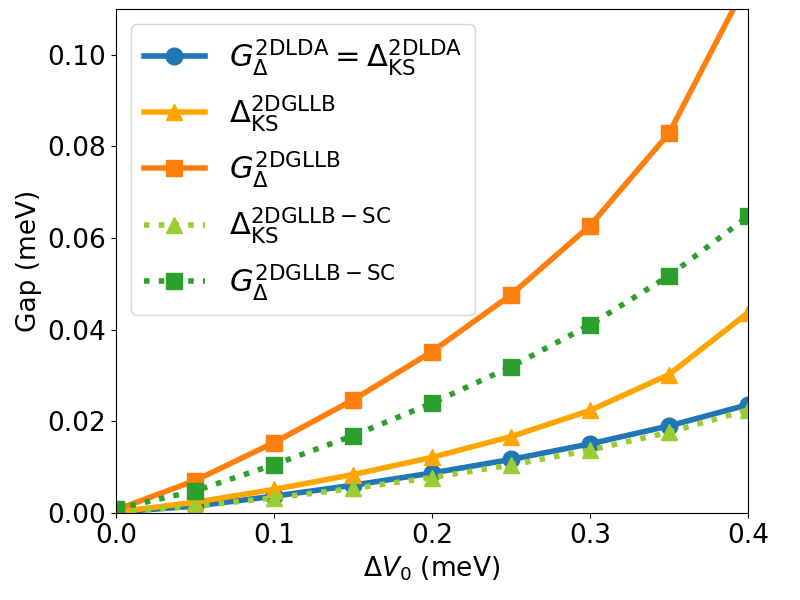}
\caption{Fundamental gap ($G_{\Delta}$) and Kohn-Sham gap ($\Delta_{\mathrm{KS}}$) at the $K$ point in artificial graphene with Kekul\'e distortion as a function of $\Delta V_0$ (see Fig.~\ref{FigPeriodicSystem}) obtained within the 2DLDA and 2DGLLB(-SC) approximations.}\label{FigKekuleBandsVSV0}
\end{center}
\vskip -0.5cm
\end{figure}

In order to examine the difference mentioned above, in Fig.~\ref{FigKekuleBandsVSV0} we plot the fundamental gaps $G_{\Delta}\uGLLBpSCp$ and $G_{\Delta}^{\mathrm{2DLDA}}$ as a function of the distortion parameter $\Delta V_0$. The differences clearly reflect the fact that
$G_{\Delta}^{\mathrm{2DLDA}} = \Delta^{\mathrm{2DLDA}}_{\mathrm{KS}}$ for all periodic 
systems as $\Delta^{\mathrm{2DLDA}}_{\mathrm{xc}}=0$. Instead, in the case of the 2DGLLB(-SC) potential, $G_{\Delta}\uGLLBpSCp \ne \Delta\uGLLB_{\mathrm{KS}}$. For vanishing distortion, we recover the Dirac cones, i.e., $G_{\Delta}\uGLLBpSCp=G_{\Delta}^{\mathrm{2DLDA}}=0$.
Fig.~\ref{FigKekuleBandsVSV0} also shows that the KS gaps are almost identical in the 2DLDA and 2DGLLB-SC cases. Both, however, differ from the case of the 2DGLLB-model potential. 

According to Sec.~\ref{SecXCv}, we expect that the  GLLB-SC results -- possibly by including proper gradient corrections which were neglected here -- should be closer to exact or close-to-exact benchmark results that could be obtained through, e.g., Hartree-Fock, GW, or equation-of-motion coupled cluster for 2D periodic systems. However, to the best of our knowledge, such benchmark results are not yet available. 

 \section{Conclusions}\label{SecConc}

In the past decades, advances in semiconductor technology have led to the development of
artificial electronic systems in reduced dimensions. In the physics of two-dimensional (2D) electronic systems we have also witnessed a shift in the focus from single quantum dots to periodic arrays of various lattice configurations; 
thus following the rapid development of atomic two-dimensional materials. 

However, the toolbox of electronic structure calculations for low-dimensional systems still requires vigorous efforts to reach the level that is currently available for regular atoms, molecules and solids in three dimensions. In this work, we have advanced a density functional approach via direct orbital-dependent modeling of Kohn-Sham potential. 
The modeling presented here is analogous to what has been successfully put forward for regular three-dimensional (3D) materials~\cite{Gritsenko1995,Kuisma2010, EJB17}. In particular, here we have worked out a shift from the domain of 3D atomic materials to the domain of 2D semiconductor devices.

Remarkably, the approach involves a computational cost which is comparable to that of a standard semi-local density functional for total energy calculations. Yet, it allows us to capture the fundamental gaps of single 2D quantum dots as well as of {\it periodic array} thereof by including key contributions which are beyond the bare Kohn-Sham gaps.

Further testing should be carried out to fully asses the quantitative aspects of the approach for 2D periodic systems. This  will require  non-trivial numerical work to implement  accurate {\it ab initio} methodologies (such as a 2D version of the GW approach or equation-of-motion coupled cluster) that --- to the best of our knowledge --- are not readily available for the systems considered in this work. In any event, the proposed model potentials are extendable to meet higher demands in accuracy. These aspects will be examined in future works.

\begin{acknowledgments}
S.~P. acknowledges financial support through MIUR PRIN Grant No. 2017RKWTMY. C. A. R. acknowledges financial support from MIUR PRIN Grant 201795SBA3.
\end{acknowledgments}

\bibliographystyle{apsrev.bst}

\end{document}